\documentclass[a4paper,twocolumn,showpacs,amsmath,amssymb,floatfix]{revtex4}
\usepackage{epsfig}
\usepackage{bm}
\usepackage{verbatim}%
\usepackage{color}
\usepackage{graphicx}
\usepackage{graphics}
\usepackage{dcolumn}
\usepackage{amsmath}
\usepackage{latexsym}
\usepackage{amsfonts}
\usepackage{amssymb}
\usepackage[rightcaption]{sidecap}
\usepackage{float}
\usepackage[utf8x]{inputenc}
\usepackage{fontenc}
\makeatletter

\newcommand{\Rmnum}[1]{\expandafter\@slowromancap\romannumeral  #1@}
\makeatother

\newcommand{\mprb}{Phys. Rev. B}

\newcommand{\be}{\begin{equation}}

\newcommand{\ee}{\end{equation}}

\renewcommand{\vr} {{\bf r}}

\def\w{\omega}
\def\e{\epsilon}
\def\vf{\varphi}
\definecolor{darkblue}{rgb}{0,0,0.4}
\definecolor{lightblue}{rgb}{0,0,0.8}

\begin{document}

\title{Some open questions in TDDFT: \\Clues from Lattice Models and Kadanoff-Baym Dynamics}
 \author{C. Verdozzi, D. Karlsson, M. Puig von Friesen, C.-O. Almbladh, and U. von Barth}
\affiliation {Mathematical Physics and European Theoretical Spectroscopy Facility (ETSF),
Lund University, 22100 Lund, Sweden }
\date{\today}

\begin{abstract}
Two aspects of TDDFT, the linear response approach and the adiabatic local density approximation,
are examined from the perspective of lattice models. To this end, we review the
DFT formulations on the lattice and give a concise presentation of the time-dependent Kadanoff-Baym equations,
used to asses the limitations of the adiabatic approximation in TDDFT. We present results for the
density response function of the 3D homogeneous Hubbard model, and point out a drawback of the linear
response scheme based on the linearized Sham-Schl\"uter equation. We then suggest a prescription on how
to amend it. Finally, we analyze the time evolution of the density in a small cubic cluster, and compare exact, adiabatic-TDDFT and Kadanoff-Baym-Equations densities. Our 
results show that non-perturbative (in the interaction) adiabatic potentials can perform quite
well for slow perturbations but that, for faster external fields, memory effects, as already present in simple
many-body approximations, are clearly required. 
\end{abstract}


\maketitle
\section{Scope of this work}\label{Intro}
A detailed knowledge and manipulation of non-equilibrium phenomena is one of the great challenges of current 
research in condensed matter - both from a fundamental point of view and from the perspective of possible applications. Theoretical 
research can contribute significantly to this endeavor, by answering a number of  important  questions. This
is what motivates the strong effort in developing theoretical methods to describe systems out of equilibrium.

In this work we focus on one of these methods, namely time-dependent density-functional theory (TDDFT) \cite{TDDFTbook}. 
TDDFT provides  the extension to the time-dependent case of static density-functional theory (DFT) \cite{HK64,KS65}. 
Its conceptual foundations were laid in the mid-eighties with the Runge-Gross theorem \cite{Hardy1}. 
Since then, alternative proofs of such theorem have been presented \cite{rvlPRL1999,ruggenthaler} (see also \cite{maitra}), 
together with variants for ensembles \cite{ensembRG}, multicomponent systems \cite{elnucRG86}, open systems \cite{diventra},
superconductivity \cite{superRG1,superRG2}, or for the case of time-dependent current-density-functional 
theory (TDCDFT) \cite{currentGhosh,vignaleRapid2004, GSEPMC10, Tokatly11,KurthStefanucci2011}.  
Furthermore, connections with other non-equilibrium methods have also been examined to clarify general 
conceptual issues in TDDFT \cite{rvl98,LundTDDFTbook}.

In TDDFT, the key variable is the one-particle density $n$, and a central ingredient is the 
time-dependent exchange-correlation (XC) potential $v_{xc}$, which incorporates all the complexities of the 
time-dependent many-body dynamics. As a result of this contracted description, and since time enters explicitly into 
the formulation, $v_{xc}$ depends in a highly non-trivial way on the entire history of the density $n$ (memory effects).
TDDFT is, in principle, an exact scheme. In practice, a major disadvantage
in concrete applications is the lack of an accurate description of dynamical inter-particle correlations.  

An easy but inadequate way to proceed is to make use of the so-called adiabatic local density
approximation (ALDA) \cite{Soven}, where the XC potential at every particular time only depends 
on the local density. This amounts to neglecting non-locality in space and memory effects in $v_{xc}$. In general, in order to
improve the ALDA, it becomes necessary to consider the issue of ultranonlocality \cite{ultra1,ultra2, ultra3}.
This means that the introduction of non-locality in time requires the inclusion of strong non-local effects
in space.

Some simplifications can be made in the case of weak perturbing fields, where linear response arguments 
apply \cite{fxc}. However, also in this case, the inclusion of non-local effects is a far-from-trivial task, as 
further discussed in the rest of the paper. Being a direct gateway to the study of important properties such as, {\it e.g.}, 
the optical response of materials, linear response within TDDFT has received a great deal of attention \cite{Botti}. 
The same applies to the problem of  improving on the ALDA away from the linear regime:
there has been a considerable theoretical effort in developing reliable XC potentials for far-from-equilibrium 
dynamics \cite{dobson,VUC97,MBW02,UT206,Toka2007,kummel1,kummel2,kummel3,baeradia,Pankratov,
Helbig2011,Fuks2011}, where memory effects are important.

In this work we analyze these two issues from a quite specific perspective, i.e. by considering TDDFT 
applied to lattice model systems. Lattice models have a long history in condensed matter physics, due to 
their conceptual simplicity and to the fact that many of such systems often admit an exact analytic solution. 
It thus does not come as a surprise that lattice models have also been investigated in the context of DFT 
and TDDFT in order to provide theoretical insight into fundamental aspects  of these frameworks.

The plan of our paper is as follows. We start in Sect. \ref{LinResp} with a presentation of the linear 
response formalism, followed by a review of (TD)DFT for lattice systems in Sect. \ref{DFTHUB}. This, 
especially for the time-dependent case \cite{Baer08,Verdozzi08}, is a rather new topic, and we will 
also survey the recent literature in the field. Then, in Sect. \ref{KBE} we proceed 
to a short summary of another method for treating non-equilibrium problems - the solving of the
Kadanoff-Baym equations (KBE) \cite{KB-book,Keldysh}. Here approximations for electron correlations
are constructed from standard methods of many-body perturbation theory (MBPT). Results from this 
approach will be used as benchmark  in our analysis of the ALDA for lattice models, which is done in Sect. \ref{RESULTS}.
In this section, we will also illustrate, using small model linear chains, some of the general aspects
of linear response theory within TDDFT, as discussed in Sect. \ref{LinResp}.
Finally, our conclusions are provided in Sect. \ref{CONCLUDE}.
\section{Linear response}\label{LinResp}
First-order perturbation theory and, what amounts to something very similar
in spirit, the theory of linear response has been around since the advent of
quantum mechanics. 
Unfortunately, first-order perturbation theory is rarely sufficient
and more and more complicated higher order corrections must be included. In
infinite systems summations to infinite order of carefully selected
corrections must be carried out. Perturbation expansions do, in principle,
not converge. They are, at best, asymptotic in character and physical
intuition reigns - even in finite systems as has been realized in recent
years \cite{Jeppe}.
Yet, perturbation theory often provides valuable insight into the
qualitative deviations between the behavior of simple models and the
realistic systems they are intended to describe.

Another very important use of perturbation theory is the description of weak
experimental probes applied to systems for the purpose of investigating
their properties without seriously affecting those very properties. A typical
example is the forced interaction between matter and weak electromagnetic
radiation. If the intensity of the radiation is chosen weak enough,
first-order perturbation theory suffices to map out the experimental results
which are then described in terms of so called linear response functions
like, {\it e.g.}, the density-density or the current-current 
response functions. 

In this section we will be mainly concerned with the density-density
response function which describes the first-order response of the system to
a perturbing local potential. The excitation of particle-hole pairs in
atoms, molecules, and solids as well as the absorption of light by such
systems are well within the realm of this theory. 

The traditional way of constructing the density response function of an
electronic system is to include the long-range Coulomb interaction through
infinite-order perturbation theory. A very successful way of accounting for
the important interaction between the excited electron and the associated
hole it leaves behind is the Bethe-Salpeter method \cite{BS,GLA}. 
Here, the vertex
corrections are obtained by solving a two-particle Schr\"{o}dinger-like
equation for the particle and the hole interacting through a statically
screened Coulomb interaction. This method has been very successful in
describing the strong excitonic effects in, {\it e.g.}, rare-gas systems
\cite{GLA}.
Unfortunately, the method is computationaly very demanding and the treatment
of systems with low symmetry like surfaces or large molecules are often
beyond reach.

In the last decade, time-dependent density-functional theory
\cite{Hardy1,Peuk}
has emerged as a tool for calculating the density responses of realistic
systems. The great computational advantage of TDDFT is that the vertex
corrections are much simpler two-point correlation functions. The drawback,
however, is the lack of a systematic approach to finding successively better
approximations to the exchange-correlation kernel which contains all the
important particle-hole interactions beyond the random phase approximation
(RPA). Within TDDFT the full density response function $\chi$ is given by 
\cite{fxc}
\begin{equation}
\chi =\chi _{0}+\chi _{0}(v+f_{xc})\chi
\end{equation}%
where $\chi _{0}$ is the non-interacting density response function expressed
in Kohn-Sham one-electron orbitals, $v$ is the bare Coulomb interaction $%
(1/r)$, and $f_{xc}$ is the famous exchange-correlation kernel into which
all the beyond-RPA many-body effects have been deferred. The kernel $f_{xc}$
is formally defined to be the second functional derivative of the XC part of
the action functional of TDDFT. A common formal starting point for obtaining
approximations to this kernel is the so called Sham-Schl\"{u}ter (SS)
\cite{SSE}
equation expressing the fact that the densities of the fully interacting
system and the Kohn-Sham system must be the same. If $G$ is the exact
one-electron Green's function of the interacting system and $G_0$ is the
non-interacting Green's function of the equivalent Kohn-Sham system these
Green's functions are connected through a Dyson-like equation
\begin{equation}
G = G_0 + G_0 ( \Sigma[G] - v_{xc} ) G
\end{equation}
where $\Sigma$ is the self-energy of the interacting system and $v_{xc}$ is
the XC part of the local one-electron potential which generates $G_0$. Since
the particle density of the system is given by the trace of the Green's
function, taking the trace of both sides of the Dyson's equation gives the SS:
\begin{equation}
\mathrm{Tr}\, [ G G_0 v_{xc} ] = \mathrm{Tr}\, [ G \Sigma [G] G_0 ]
\end{equation}
The SS is thus an equation for the determination of the exact $%
v_{xc}$ of TDDFT. A further variation with respect to, {\it e.g.}, the external
potential provides an expression for the XC kernel $f_{xc} = \delta
v_{xc}/\delta n$ although it is rather complicated and has so far, to our
knowledge, not been used by anyone for practical calculations.

By doing ordinary diagramatic perturbation expansions for the Green's function 
$G$ and the corresponding self-energy $\Sigma$ in the SS, this equation
provides a natural connection between TDDFT and many-body perturbation theory. 
Along these lines, a diagrammatic procedure for finding
better kernels $f_{xc}$:s was developed in Ref. \cite{mbptfxc},  but it still remains untested.

What has been used in several applications is a simplified version of the SS
called the linearized Sham-Schl\"{u}ter equation (LSS). This is obtained by
replacing every interacting Green's function $G$ by the non-interacting
Kohn-Sham Green's function $G_0$ in the SS. The resulting equation
\begin{equation}
\mathrm{Tr}\,[G_{0}G_{0}v_{xc}]=\mathrm{Tr}\,[G_{0}\Sigma \lbrack
G_{0}]G_{0}]
\end{equation}%
can also be derived from the variational many-body approach \cite{ABL}.
One then starts from the so called Klein \cite{klein} functional for
the total action. Subsequently, one chooses to restrict the normally free
variations of the one-electron Green's function $G$ to those non-interacting
ones that can be generated by multiplicative, local one-body potentials.
This demonstrates that the LSS is a more accurate expression for generating
XC potentials of TDDFT than is suggested by its original derivation where it
was simply the lowest order result in a perturbation expansion. We notice
that, within the LSS approach, very high-order correlation effects can be
accounted for by choosing a very sophisticated self-energy. We should,
however, also point out that there is an inherent lack of self-consistency in
the LSS approach. The Kohn-Sham Green's function $G_{0}$ is certainly not
generated by that Dyson equation which has $\Sigma \lbrack G_{0}]$ as a
self-energy. This lack of self-consistency could easily lead to violations of
some of the conservation laws and consistency requirements which one usually
takes for granted within variational formulations.

Unfortunately, there is a much more severe problem associated with use of
the LSS approach as we will now discuss. One can easily perform a variation
with respect to the externally applied potential in the LSS. Containing,
however, only the non-interacting Green's function $G_0$ this is equivalent to
performing a variation $\delta V$ in the full effective Kohn-Sham potential $%
V$. Using the obvious fact that
\begin{equation}
\delta G_0 = G_0 \delta V G_0
\end{equation}
such a variation leads to the following equation for the XC kernel $f_{xc}$
\begin{align}
&\!\!\!\int d2d2^{\prime} \,\,\chi_0(1,2) f_{xc}(2,2^{\prime}) \chi_0(2^{\prime},1^{\prime}) =\nonumber\\
&\!\!\!-i \int d2d3d4d5 \,\, G_0(1,2) G_0(3,1) \frac{\delta \Sigma (2,3)}{\delta G(4,5)}
G_0(1^{\prime},5) G_0(4,1^{\prime}) \nonumber\\
&\!\!\!-i \int d2d2^{\prime} \,\, G_0(1,1^{\prime}) G_0(1^{\prime},2) \Delta(2,2^{\prime}) G_0(2^{\prime},1)
\nonumber\\
&\!\!\!-i \int d2d2^{\prime} \,\, G_0(1,2) \Delta(2,2^{\prime}) G_0(2^{\prime},1^{\prime}) G_0(1^{\prime},1)
\end{align}
where the quantity $\Delta$ is the difference between the chosen self-energy 
$\Sigma$ and the XC potential $v_{xc}$
\begin{equation}
\Delta (1,1^{\prime}) = \Sigma(1,1^{\prime}) - v_{xc}(1) \delta(1,1^{\prime})
\end{equation}
If the four-point vertex function $\delta \Sigma /\delta G$ is taken from
Hartree-Fock theory, which means that it becomes the bare Coulomb
interaction, the resulting expression for the XC kernel $f_{xc}$ is called
the exact-exhange approximation (EXX) within TDDFT. In the course of time,
this approximation has been derived by many researchers and it has been
given many different names like the optimized potential method (OPM), the
optimized effective potential (OEP) approach, or the exchange-only
approximation (EOA). It has been shown to have many nice properties in
extended systems and it gives very accurate total energies in both finite
and infinite systems ranging from atoms and molecules to the electron gas.
The spectral properties of the resulting density response function was
rather recently investigated at length in series of papers \cite{KvB,MHvB07,MHvB09}. 

In these works, it was discovered that the resulting response function of finite
systems actually has poles in the upper half plane thus precluding a
resonable description of the optical response of the system at higher
energies. This problem is associated with rather unexpected zeroes of the
non-interacting response function $\chi_{0}$ at certain frequencies. The
presence of such zeroes was, however, established a long time ago \cite{MK}.
From the defining equation above for the kernel $f_{xc}$ we
see that one has to invert the response function $\chi_{0}$ twice in order
to obtain $f_{xc}$. Thus, a zero in $\chi_{0}$ produces a double pole in
$f_{xc}$ which, in turn, causes the full response function $\chi $ to have an
unphysical pole in the upper half plane. It is important to notice that this
problem has nothing to do with the degree of sophistication by which one
tries to incorporate the correlation effects. Any choice of self-energy will
produce the same unphysical result. 
\section{DFT and TDDFT for lattice models: the case of the Hubbard Model} \label{DFTHUB}
\noindent The model of specific interest to the present paper is the Hubbard model \cite{Hubbard}. This lattice system is one of
the most studied in research on strongly correlated systems, and  is also the one which has received most attention 
in the context  of static and time-dependent DFT studies. For the non-equilibrium case, it is described by the Hamiltonian
\begin{equation}
H\!= -t\!\!\sum_{\langle RR'\rangle \sigma }a^{+}_{R\sigma}a_{R'\sigma}\!\!+\sum_{R}U_R\hat{n}_{R\uparrow}\hat{n}_{R\downarrow}
+ \sum_{R\sigma} w_{R\sigma} (\tau)  \hat{n}_{R\sigma}.
\label{Hamiltonian}
\end{equation}
Eq. (\ref{Hamiltonian}) is a direct generalization of the usual Hubbard Hamiltonian
to the inhomogeneous, time-dependent case, and in presence of spin-dependent potentials \cite{t_meaning}. Such a Hamiltonian offers one 
of the simplest descriptions of the competing behavior between the itinerant and localized behavior of electrons 
in the presence of interactions. 
In Eq. (\ref{Hamiltonian}), the term $\hat{W}(\tau)\equiv  \sum_{R\sigma}  w_{R\sigma} (\tau)  \hat{n}_{R\sigma}$ 
describes a local (in space and time), spin- and/or  time-dependent potential ($\tau$ denotes the time variable).
For convenience, in $\hat{W}(\tau)$ we 
separate the static and time-dependent parts: $w_{R\sigma} = \epsilon_{R\sigma} + v_{R\sigma} (\tau)$. In the static case,
all $v_{R\sigma}(\tau)=0$; furthermore, in spin-independent  formulations, $\epsilon_{R\sigma}$ and
$v_{R\sigma}(\tau)$ are independent of $\sigma$. Finally, the standard single-band homogeneous Hubbard 
Hamiltonian \cite{Hubbard} is recovered when $U_R=U$ and $w_{R\sigma}=0$. 

\subsection{General aspects of ground state lattice DFT}
The use of the Hubbard model in the context of static DFT was introduced 
in a comparative study of many-body and DFT Fermi surfaces \cite{GSDFT2}.
However, a DFT formulation based on the local lattice occupation numbers $n_R$
had already been introduced by the same authors in earlier work  \cite{GSDFT1,GSDFT1bis}, to study
the XC discontinuity in a semiconductor model. Later on, a more general formulation of static lattice DFT
was presented \cite{GSN}, and an analysis of the local density approximation (LDA) was
performed for the 1D Hubbard model (where an exact solution based on Bethe-Ansatz is possible \cite{LiebWu}).
At the same time, other formulations were proposed, which use the lattice one-particle density matrix 
$\gamma_{RR'}$ as the basic variable \cite{Godby,Pastor}. Further significant progress within static lattice DFT
was made in Ref. \cite{LimaPRL03}. In this work, a LDA based on the Bethe-Ansatz
solution (henceforth denoted BALDA) for the $v_{xc}$  was proposed, 
suitable for a DFT of the inhomogeneous 1D Hubbard model. An analytical paramatrization of the XC energy and
potential was also provided. In a series of works 
\cite{LimaPRL03, LimaEPL02, CapellePRB, CapellePolini, FrancaCapelle}, the BALDA for $v_{xc}$ was tested 
and benchmarked against exact diagonalization, density-matrix renormalization group (DMRG)
and quantum Monte Carlo calculations and shown to attain an accuracy of the order of a few percent for energies,
particle densities and entropies.

In this paper, we do not consider magnetic effects, and limit ourselves to a discussion of a spin-independent DFT 
and TDDFT for the Hubbard model. In static DFT, and in standard notation, we can write for the ground-state total energy \cite{GSDFT2,GSN}:
\begin{align}
E_v [n] \equiv T_0[n] + E_H[n] + E_{xc}[n]+\sum_i v_{ext}(i) n_i ,
\label{theory::Exc}
\end{align}
where $v_{ext}$ is the static external field (in the notation of Eq. (\ref{Hamiltonian}),
$v_{ext}(i)\equiv \epsilon_i$). In Eq. (\ref{theory::Exc}),
$n_i=\sum_\sigma n_{i\sigma}$,  while $T_0[n]$ and $E_H = \frac{1}{4} \sum_i U_i n_i^2$ are, respectively, 
the non-interacting kinetic energy and the Hartree energy.
To perform a local density approximation,  $E_{xc}$ is obtained from a homogeneous Hubbard model, chosen as a suitable reference system:
\be
E_{xc} = E - T_0- E_H .
\ee
To obtain the  exchange-correlation potential $v_{xc}$, one performs the derivative of the XC energy/site $e_{xc}\equiv E_{xc}/L$
with respect to the density (in the general case,  
a functional derivative should be considered):
\be
v_{xc}=\frac{\partial  e_{xc}(n,U)}{\partial n}.
\ee
Due to electron-hole symmetry,
\be
e_{xc}(n,U)= e_{xc}(2-n,U),
\ee
in the entire density range $[0,2]$; thus  
\be
v_{xc}(n)=-v_{xc}(2-n).
\ee
Finally, a local density approximation is performed: 
\be
v_{xc}(i) = v_{xc} (n_i).
\ee
To ease the numerics when $v_{xc}$ is discontinuous (see below), one can use a slightly smoothened version of $v_{xc}$ 
near $n=1$. 
The $v_{xc}$ thus obtained can be used in ground-state DFT-LDA calculations, which amounts to solving
self-consistently the Kohn-Sham (KS) equations
\begin{equation}
( \hat{t} + \hat{v}_{KS} ) \varphi _\kappa = \varepsilon_\kappa\varphi _\kappa \ \ ,    \label{staticKS}                             
\end{equation}
 where $\hat{t}$ denotes the matrix for the single-particle hoppings  among nearest-neighbor sites, 
 and $\varphi_\kappa$
is the $\kappa$-th single-particle KS orbital, with $n_i = \sum_{\kappa \in occ} |\varphi _\kappa (i)|^2$.
The effective potential matrix is diagonal: $(\hat{v}_{KS})_{ii}=v_{KS} (i) = v_H(i) + v_{xc} (i) + v_{ext}(i)$, with  $v_H(i)=\frac{1}{2} U_i n_i$
being the Hartree potential.


\subsection{DFT for the Hubbard model and dimensionality}\label{DFTHub}

{\it 1D Hubbard model.-}
At half filling, i.e. when $n=1$, for the infinite, homogeneous Hubbard model, the exact ground-state energy (obtained by the Bethe-Ansatz) is \cite{LiebWu}:
\be
e(1,U)=-\frac{2\beta}{\pi}\sin(\frac{\pi}{\beta})=\!\int_{0}^{\infty} \!\!\!dx\frac{-4 J_0(x)J_1(x)}{x[1+exp(Ux/2)]}.\label{1Dn=1}
\ee
The interpolation formula proposed for the XC energy/site at general densities is \cite{LimaPRL03}
\be
e_{xc}(n,U)= -\frac{2\beta(U)}{\pi}\sin(\frac{\pi n}{\beta(U)}) + \frac{4}{\pi}\sin(\frac{\pi n}{2})-\frac{Un^2}{4},
\ee
where $\beta$ depends on $U$ but not on the density, and is determined by Eq. (\ref{1Dn=1}).
It is easily seen that $\beta(U=0)=2$, and $\beta(U \rightarrow \infty)=1$.  
This gives for $v_{xc}$: 
\be
v^{n<1}_{xc}(n) = -2\cos(\frac{\pi n}{\beta(U)})+2\cos(\frac{\pi n}{2})-\frac{Un}{2} \label{LSOC}
\ee
and $v^{n>1}_{xc}(n) = - v_{xc}(2-n)$. 
With the parametrization of Eq. (\ref{LSOC}) for $v_{xc}$, the discontinuity of $v_{xc}$ at half-filling
is $\Delta v_{xc}=4\cos(\pi/\beta(U))+U$ \cite{LimaEPL02}. This expression recovers correctly the exact asymptotic limits 
$\Delta^{U\rightarrow 0}= (8/\pi)\sqrt{U}e^{-2\pi/U}$ and  $\Delta^{U\rightarrow \infty}=U-4+8\ln(2)/U $.
For $U\gtrsim 2$ the parametrization of Eq.(\ref{LSOC}) agrees within few percent with the full numerical
Bethe-Ansatz solution. For $U\lesssim 2$,  Eq. (\ref{LSOC}) is not equally accurate. We finally mention that, 
very recently, a parametrization of the XC potential has become available also for the 
spin-dependent case \cite{SDBALDA}.
\newline
{\it 2D Hubbard model.-}
In 2D, the Hubbard model has been investigated via DFT on the graphene lattice.  In this case,
the expression for the XC energy/site is \cite{Harju}
\be
e_{xc}(n,U)=a(e^{-b U^2}-1)e^{-(c |n-1|-d)^2}.
\ee
The parameters $a,b,c,d$ were determined by a fitting procedure to the ground-state energy calculations, performed
with the exact diagonalization method. \newline

{\it 3D Hubbard model.-}
To date, the only case considered in the literature is that of the simple cubic lattice \cite{DKAPCV11},
where the ground-state energy of the uniform system was computed within dynamical mean field fheory (DMFT)
\cite{metznervollhardt,revdmft}.  The short summary below follows rather closely the one originally given in \cite{DKAPCV11}.

Initially devised for lattice models with infinite connectivity (where the self-energy $\Sigma$ 
is local in space), DMFT deals non-perturbatively with correlations; in 3D, it can be seen as an approximation scheme with 
a local self-energy. 
DMFT maps a Hubbard model on a simple 
cubic lattice onto a local problem representing one of the lattice sites (site 0)
hybridized with a bath (the rest of the lattice) \cite{revdmft}. 
A Hamiltonian description of the local problem can be recovered, by identifying the site $0$ 
with the impurity site of an Anderson impurity model (AIM), and introducing auxiliary 
degrees of freedom $\{a^{\dagger}_{k\sigma}\}$:
\begin{equation}
\label{aim}
\!{\cal{ H}}_{AIM}  = {\cal{H}}_{imp}+\sum_{k,\sigma}
\left[ \epsilon_k a^{\dag}_{k\sigma} \, a_{k\sigma} + V_{0k}\,
(a^{\dag}_{k\sigma} c_{0\sigma} + \mbox{h.c.}) \right],
\end{equation}
where $ {\cal{H}}_{imp} = Un_{0\uparrow} n_{0\downarrow} - \mu (n_{0\uparrow} +n_{0\downarrow} )$.
The value of the parameters  $\{\epsilon_k, V_{0k}\}$ in the auxiliary Hamiltonian ${\cal{H}}_{AIM}$
are determined self-consistently,
i.e. when the impurity single-particle Green's function $G(i\omega_n)$  in Matsubara space \cite{definitions}
becomes identical to the local
lattice Green's function with identical self-energy $\Sigma(i\omega_n)$:
\begin{equation}
 G(i \omega_n) = \int d\epsilon  D(\epsilon)[i\omega_n+\mu-\epsilon-\Sigma(i\omega_n)]^{-1}.
\end{equation}
Here, $\Sigma(i\omega_n)= \mathcal{G}_0^{-1}(i\omega_n)-G^{-1}(i\omega_n)$
is the local Dyson equation,   
$\mathcal{G}_0^{-1}(i\omega_n)=i\omega_n+\mu-\sum_k \frac{V_k^2}{i\omega_n-\epsilon_k}$,
and
$D(\epsilon)$ is the non-interacting lattice density of states.
Once at self-consistency, the relevant quantities ({\it e.g.}, the double occupancy $\langle n_{0\uparrow} n_{0\downarrow} \rangle$,
the total energy, etc.) are finally computed. It is perhaps worth mentioning that schemes other than
DMFT could be used  to determine the $e_{xc}$ of the homogenous system,
{\it e.g.} the Quantum Monte Carlo method or the Gutzwiller approximation (such work is currently under way).

\begin{figure}[t]
\includegraphics[width=85mm,angle=-0]{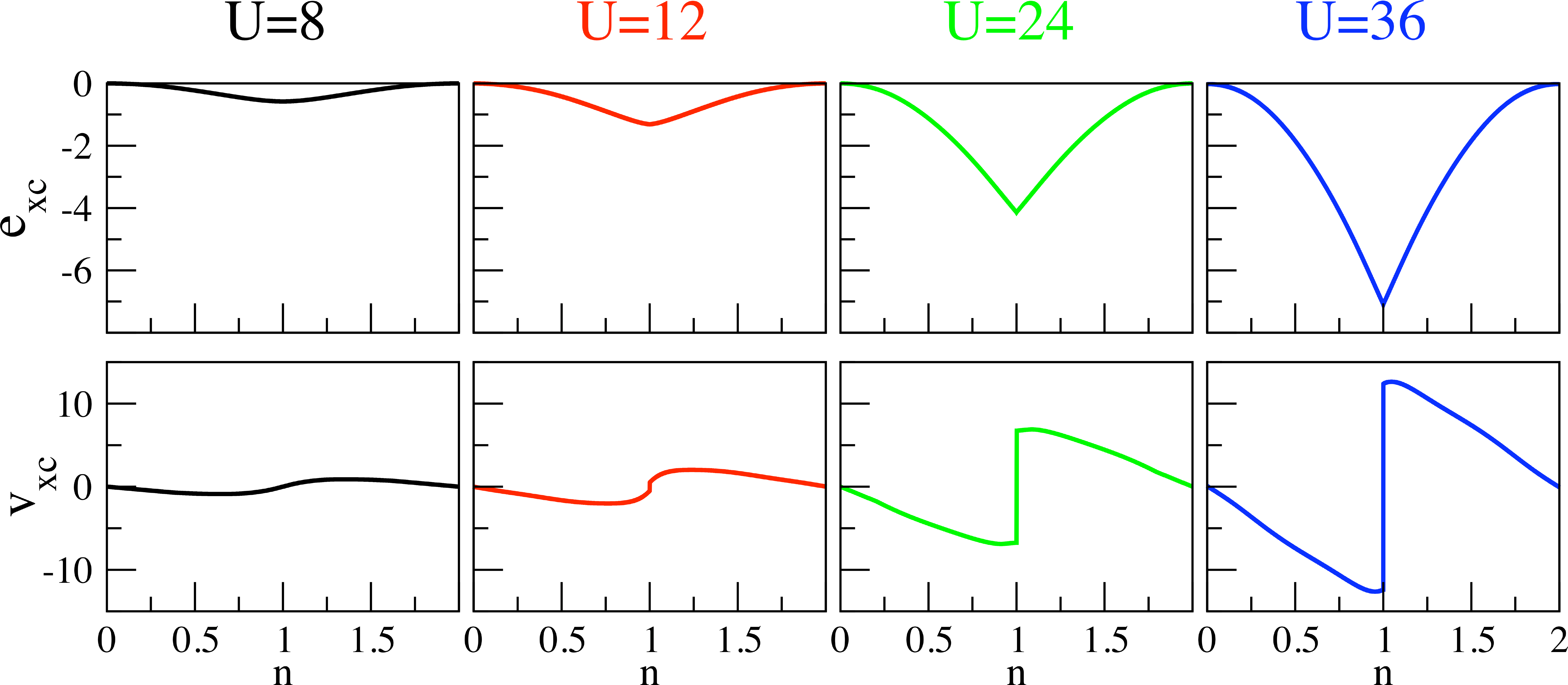}
\caption{(Color online) Exchange-correlation energies $e_{xc}$ (top) and potentials $v_{xc}$
(bottom) for the homogeneous 3D Hubbard model, as a function of the density, for several values of the interaction $U$.}
\label{vxcDMFT} 
\end{figure}

In Fig. \ref{vxcDMFT} we present results for $e^{DMFT}_{xc}$ and $v_{xc}^{DMFT}$ for several $U$ values. 
On increasing $U$, the curvature of $e_{xc}$ changes, and for
$n \approx 1$  a cusp develops above a critical
value $U_c^{Mott}$. As a consequence, a discontinuity shows up in $v_{xc}$. This
latter feature is the manifestation of the Mott-Hubbard metal-insulator transition 
within a DFT description. This behavior is rather different from what is observed in the 1D Hubbard model, 
where $v_{xc}$ is discontinuous for any $U>0$  \cite{LimaPRL03}. The XC potentials shown in Fig. \ref{vxcDMFT}
will be utilized in Sect. \ref{RESULTS}.

\subsection{Lattice TDDFT}
We have seen that  DFT can be a viable route to the description of the ground-state properties
of strongly  correlated systems such as the Hubbard model. A subsequent step
would be to look at the dynamical properties of the Hubbard model via TDDFT.
Similarly to the case of realistic materials, this could be done at two different levels: 
1) One can invoke a linear response treatment to external fields, using 
the apparatus of many-body perturbation theory,
starting from the Kohn-Sham Hamiltonian; 
2) In presence of strong time-dependent fields, one could instead resort to a full dynamical description based 
on TDDFT, by propagating in time the KS equations.

For 1D Hubbard-type Hamiltonians, initial work in the first direction was made in Refs. \cite{Arya1, Arya2, Magyar}. 
In these papers, the main emphasis was on
the study of the Hubbard model as a simplified system to gain insight into the general aspects of TDDFT, 
specifically the kernel $f_{xc}$, and no real-time dynamics was performed. TDDFT in the linear response regime
was also applied to a 1D spinless fermion model with nearest-neighbour interactions \cite{Schenk08}, where
comparisons between exact results and an LDA based on the Bethe-Ansatz were reported. 
Less is known for $D>1$, and we will present below new results for 
the linear response of the 3D Hubbard model within the  TDDFT-ALDA.

A TDDFT approach to the real-time dynamics of the Hubbard model out of equilibrium 
was initially considered in \cite{Verdozzi08}. In this work, the exact many-body 
non-equilibrium dynamics of small 1D Hubbard chains of different length 
and particle density was performed, and from it, via a reverse-engineering procedure, the exact, 
time-dependent XC potential was obtained, by propagating the time-dependent Kohn-Sham equations:
\be
\left( \hat{t} + \hat{v}_{KS}(\tau)\right ) \varphi_\kappa(\tau) = i \partial_\tau  \varphi _\kappa(\tau).\label{TD_KS}                                
\ee
Eq. (\ref{TD_KS}) is the generalization of Eq. (\ref{staticKS}) to the time-dependent case.
In \cite{Verdozzi08}, exact results for the density and the XC potential were compared to those from
an approximate $v_{xc}$ obtained within the ALDA, and an analysis of non-local adiabatic 
effects in $v_{xc}$ was carried out. For the ALDA, in analogy to the static case,
the 1D Hubbard model was chosen as the reference systems, so that 
\be
v^{ALDA}_{xc}(R,t)\equiv v^{BALDA}_{xc}(n_R(t)), 
\ee
where $v^{BALDA}_{xc}(n)$ is given by Eq. (\ref{LSOC}).
In Ref. \cite{Verdozzi08}, the treatment was limited to spin-compensated systems,
and the spin-dependent case was presented in \cite{Polini08}, where TDDFT results were
compared to those obtained with the time-dependent DMRG method.

An early rigorous discussion of the formal aspects of TDDFT can be found in 
\cite{Baer08}, where an example of particle density which is non $v_0$-representable
on the lattice was provided. In \cite{Baer08} it was shown that a tight-binding Hamiltonian may in general 
be unable to reproduce a specific given density. The case considered was that of a dimer with 
one orbital/site. The issue of $v_0$-representability was further analyzed in \cite{Ullrich08, Verdozzi08}. 
As shown in \cite{Ullrich08}, for a 1D lattice with $L$ sites, the condition $|S_k|\le 2\sqrt{n_k n_{k+1}}$,
with $S_k=\sum_{i=1}^k \dot{n}_i$ (this expresses current 
conservation), is necessary and sufficient for $v_0$-representability for $N=1$ particle.
In a lattice Hamiltonian, the spread of the eigenmodes scales with the hopping term $t$.
 According to the time-energy uncertainty relation, the range of possible response times scales
also with $|t|$.  Thus, the larger the hopping, the faster the system can respond in time, and the broader 
the range of $v_0$-representability in time \cite{Ullrich08}. The above condition  for $S_k$ is
only a necessary one for $N>1$ \cite{Verdozzi08}, also when Hubbard interactions are taken into account. 
For the special case of a dimer, and using the Cauchy-Schwartz inequality, one can show that the exact 
density of a Hubbard dimer is $v_0$-representable \cite{Verdozzi08}.
\subsection{Developments in lattice DFT and TDDFT, and applications.}
\noindent {\it $\bullet$ Electric polarizability.-} Very recently, lattice DFT has been used to determine the polarizability $\alpha$ of the 1D Hubbard model \cite{AkandeSanvito10}. Results for $\alpha$ from lattice DFT are in good agreement with DMRG ones. 
In particular,  the response of the XC potential is in the
same direction of the perturbing potential.  Also, the possibility of dealing in lattice DFT with large samples 
made possible to examine the scaling properties of $\alpha$ \cite{AkandeSanvito10}.\newline
{\it $\bullet$ Magnetic properties.-} lattice DFT has also been used to study the Hubbard model on the graphene lattice, where it correctly
describes the ground-state spin configuration of large graphene clusters (flakes) \cite{Harju}. \newline
{\it $\bullet$ Quantum Information and Entanglement entropy.-}  A key quantity to quantum information phenomena
is the entanglement, or quantum correlations, of a system. Entanglement has also been studied in the
context of quantum phase transitions. A lattice DFT approach to the entanglement entropy of the Hubbard
model was presented in \cite{FrancaCapelle} and, very recently, an explicit density functional for the 
entanglement was also provided \cite{FrancaDamico11}. An TDDFT-ALDA approach to the time evolution of
entanglement entropy was used in \cite{KarlssonEPL2011}, to discuss the expansion of ultracold fermion 
clouds in optical lattices. \newline
{\it $\bullet$ Quantum transport.-} 
A combination of DMRG and lattice DFT was used to gain insight into the exact ground-state XC functionals for a correlated-electron model system coupled to external reservoirs \cite{SchmittEver08}. The specific model considered was the so-called interacting resonant level model, for which DMRG and DFT conductances were found in good agreement. Very recently, a comparative assessment of lattice DFT and DMRG in transport has been provided in \cite{schenk2011},
while an application of the spin-dependent BALDA \cite{SDBALDA} to quantum transport can be found in \cite{Thijssen}.

The role of a discontinuity in a
time-dependent description of quantum transport has been recently examined within lattice TDDFT-ALDA \cite{KSKVG10}. Following the time-evolution of a single Anderson impurity attached to two biased leads, a dynamical notion of the Coulomb blockade was provided. Accordingly, Coulomb blockade manifests itself as a
periodic sequence of charging and discharging of the nanostructure. This emerges also from 
a description based on TDCDFT \cite{KurthStefanucci2011}. In passing,
we mention that ground-state current DFT has been also applied to lattice models \cite{Schenk2, AkandeSanvito11}.
For 1D Hubbard rings threaded by  a magnetic flux,  the effects of 
lattice impurities on the persistent currents and on the Drude weights has also been examined \cite{AkandeSanvito11}. \newline
{\it $\bullet$ Connection to many-body approximations, and critical analysis of  the local density approximations.-}
The so-called variational approach to TDDFT \cite{{ABL},{varfxc}} has the advantage of a systematic
inclusion of many-body contributions in the XC potential. In this way non-locality
in space and memory effects can be properly included, once  $v_{xc}$
is retrieved from the many-body self-energy via the time-dependent Sham-Schl\"uter equation \cite{RobertShamSh96}.
This well established relation between many-body perturbation theory and TDDFT on the Keldysh contour was numerically examined in \cite{pva2}, by looking at exchange-correlation potentials obtained via time-dependent reverse engineering using the time-dependent densities from  the KBE. The performance of two self-interaction correction schemes for the 1D Hubbard model has been scrutinized in \cite {CapelleSIC}, while 
a shortcoming of the BALDA was pointed out in \cite{Capelle2kf4kf}, specifically its capability to
reproduce correctly the Friedel oscillations in the inhomogeneous 1D Hubbard model.
Finally, the lattice DFT for the 1D Hubbard model has been used to illustrate the Hohenberg-Kohn
mapping between densities and ground-state wave functions in terms of the
metric-space aspects of the Hilbert space \cite{Capelle_Metric}. \newline
{\it $\bullet$ Cold atoms and dynamical quenches.-}  
In optical lattices \cite{opticalattices}, it is possible to study fermionic and bosonic atoms with repulsive and/or  
attractive interactions and (because an accurate tunability of the lattice parameters is possible)
often more directly and easily than in solid-state experiments. Trapped ultracold atoms 
on an optical lattice permit to study different ground-state scenarios for the Hubbard model.
Lattice DFT and lattice TDDFT have been used to investigate these systems. For example, an extensive study of the ground-state properties of trapped repulsive fermions in a 1D Hubbard lattice was provided in
\cite{XianlongPoliniTosiCampoCapelleRigol06}. Here, a comparison of lattice DFT and quantum Monte Carlo density 
profiles was performed, and an detailed microscopic picture of the consequences of the interplay between particle-particle interactions and confinement was given. 

A similar analysis within lattice DFT has been carried out in
\cite{Hu2010} for the case of a 1D Hubbard model with attractive interactions.
For lattice TDDFT, a first application to fermions on optical lattices was to compare 
(spin-dependent) ALDA and DMRG dynamics after the quench of a localized external perturbation \cite{Polini08}. 
A more recent study of the dynamics after a local quench can be found in \cite{Xianlong2010}.
The effect of a global quench in 1D, i.e. the removal of a parabolic trapping potential,
was studied in \cite{KarlssonEPL2011}, where quenches of different speed (from adiabatic to sudden)
were applied, showing a dynamical self-stabilization of the Mott insulator phase in the density profile. 
In this work, the role of entanglement as a pointer of dynamical phase changes was also analyzed,
and the system's thermalization was discussed from a TDDFT perspective. Finally, the beating and 
damping regimes of the Bloch oscillations in a 3D Hubbard optical lattice has also 
been investigated with TDDFT \cite{DKAPCV11}.
%
%
\subsection{TD[C]DFT on the lattice: Runge-Gross theorem and connections to lattice TDDFT}
There is at present no formulation of  the Runge-Gross (RG) theorem of TDDFT  for  the lattice.
The reason for this has been explicitly discussed in \cite{KurthStefanucci2011} and, in essence,
it has to do with the fact that a {\it reductio ad absurdum} strategy, along the lines of the 
RG theorem, is not possible. This is because, in the proof for the lattice, terms 
related to the one-particle density matrix appear, which do not carry a definite sign.
This observation corroborates the fact that, for lattice Hamiltonians, one can actually find 
examples which contradict the uniqueness of the correspondence between densities and potentials.

If one adopts the bond current as the basic variable, a one-to-one mapping
can be established \cite{GSEPMC10} betweeen current densities and Peierls phases \cite{Peierls,Graf} of the bond-hopping terms. 
In this way a rigorous formulation of TDCDFT on the lattice becomes possible \cite{GSEPMC10, Tokatly11, KurthStefanucci2011},
either in analogy with the original RG theorem \cite{GSEPMC10}, or via a reformulation
based on non-linear differential equations methods \cite{Tokatly11}. 

In addition to establish a rigorous connection also for the densities, a lattice TDCDFT formulation has 
the further advantage of being valid in the presence of magnetic fields of a general kind, which enter via the Peierls phases
(more precisely, the Peierls phases describe the effect of the vector potential $\bf {A}$).  
This is not the case of TDDFT; it can be easily shown that, by a local gauge transformation,
the scalar onsite potentials of TDDFT can be removed from the Hamiltonian while, at the same time, 
introducing Peierls phases $\phi_{ij}=\alpha_i-\alpha_j$ in the hopping term between sites $i$ and $j$.
The sum of phases of such specific form over a lattice plaquette adds up to zero, and corresponds
to a null line integral of ${\bf A}$ over a closed loop, i.e. to a zero magnetic flux across the plaquette.
For these cases, i.e. in the absence of magnetic fields, TDDFT could be used; otherwise, TDCDFT
should be considered. 

In conclusion, care must certainly be exerted in using lattice TDDFT when dealing with issues 
for which the uniqueness of the density-potential mapping is relevant.  
It should, however, still be possible to use lattice TDDFT on heuristic grounds,
even if, at present, the practical implications of the lack of a rigorous formulation largely remain to be seen.
\section{Kadanoff-Baym Dynamics }\label{KBE}
The main aim of the non-equilibrium Green's function (NEG) technique 
\cite{KB-book,Keldysh,Danielewicz,Bonitzbook} is to 
obtain expectation values of single-particle operators and the total energy 
of an interacting system subject to an external field.  
The NEG technique is used in a wide variety of fields, 
treating both real and model systems \cite{Jauho,Stefanucci04,Robert,
Bonitzdots,Galperin,Thygesen1,Olevano,RobertHubb,Freericks, pva1}. 
The method has the advantage of having a direct connection to 
MBPT, in which conserving \cite{KB-book} approximations of increasing
complexity can be constructed systematically and where memory
effects \cite{Danielewicz,Bonitzbook} are automatically built in.
However, when the NEG technique is used in conjunction with MBPT,
the method has also a few limitations, some of which are practical in character. 
Before discussing further this point, we proceed now to a brief summary
of the approach.
\subsection{Formalism}
The central object in the NEG echnique is the path-ordered, one-particle 
Green's function,
\begin{equation}
G(1,2)=-i\langle T_\gamma\left[\psi(1)\psi^\dagger(2)\right]\rangle, 
\label{Green}\end{equation}
where, in more detail,  1 denote single-particle space/spin and time labels, $r_1 \sigma_1 t_1$.   
Here, $T_\gamma$ orders the times $t_1,t_2$ on the Keldysh contour \cite{Keldysh}
$ \gamma$ ($t_1,t_2$ can be either real or imaginary), and the field operators 
are in the Heisenberg picture; the brackets 
$\langle\rangle$ denote averaging over the initial state (ground state or thermal 
equilibrium).
 
The $G$ is determined by the Kadanoff-Baym equations, its equations of motion.
Specializing to time $t_{1}$, we have \cite{notation}
\begin{equation}  
\left(i\partial_{t_{1}}-h\left(t_{1}\right)\right)G\left(t_{1},t_{2}\right)=
\delta(12)+\int_{\gamma}\Sigma\left(t_{1},t\right)G\left(t,t_{2}\right)dt. \label{KBEeq}
\end{equation}
Here $h$ is the single-particle Hamiltonian and $\Sigma$, 
the kernel of the integral equation, is the self-energy,
treated within a given many-body approximation (MBA). Eq. (\ref{KBEeq}) and the one corresponding to $t_2$ are solved numerically,
and for non-isolated systems, $\Sigma$ contains an additional  contribution,
the embedding self-energy $\Sigma_{emb}$, to treat the coupling system-environment \cite{Stefanucci04,RobertHubb,pva2}.

The MBA:s we consider in this paper are the second Born, the $GW$ and the 
$T$-matrix approximations (BA, GWA and TMA respectively). 
The self-energy in the BA includes all terms up to second order. 
The GWA \cite{Hedin} amounts to add up all the bubble diagrams which 
give rise to the screened interaction, $W=U+UPW$, where $P(12)=G(12)G(21)$. 
In this case the self-energy is $\Sigma(12)=G(12)W(12)$.
In a spin-dependent treatment of the TMA \cite{TMA,pva2} 
one constructs the $T$ by adding up all the ladder 
diagrams, $T=\Phi-\Phi U T$, where $\Phi(12)=G(12)G(12)$. 
The expression of the self-energy then becomes 
$\Sigma(12)=\int U(13)G(43)T(34)U(42)d34$ (for further
details see {\it e.g.} \cite{pva2}). 

All these approximations are conserving 
\cite{KB-book}, i.e. the macroscopic conservation laws 
(for the energy, the number of particles, etc.) hold, which is
a key requirement for time dynamics. The fact
that the approximations are conserving does, however, not guarantee the 
soundness of other important features, as shown in the next Section. 
\subsection{Some drawbacks of KBE+MBPT}

On the practical side, a first limitation is that the Green's
function is a two-point function and thus scales quadratically with the 
propagation time. A second one is that the KBE must be solved 
self-consistently with a non-local kernel at every point in time. Currently, this is
viable for very small systems, whilst larger ones can 
only be treated in the steady state regime (where only one time 
variable is required). 

On the more fundamental side, the NEG approach within MBPT 
manifests other undesirable traits, for example perturbation schemes
may break down for strong interactions. However, even for converging
schemes, unphysical features can emerge, which can sometimes play 
an important role. For example, in MBPT, the ground-state spectral function of a finite 
system contains infinitely many poles, whereas in the exact solution the 
number of poles in finite. Moreover, in the KBE-based time dynamics, the 
MBA:s introduce an artificial correlation induced damping which, in some cases, 
can completely dominate the evolution. For finite systems, such damped dynamics results 
in a steady (i.e. never decaying) state.
In extended systems, a similar behavior is observed, but with a difference,
namely arbitrarily long-lived states occur which, however, eventually decay into a unique
steady state. This behavior is certainly artificial in finite systems, while
its physical soundness in extended systems is much harder to establish.

Another drawback concerns KBE and correlation functions:
It is well known that single-particle Green's functions give access to
certain two-particle properties  such as, {\it e.g.}, the total energy. 
In particular, they can be used to determine the density-density correlation 
function at equal times, {\it e.g.} the so-called double occupancy, 
$\left<\hat{n}_{R\uparrow}\hat{n}_{R\downarrow}\right>$ 
\cite{pva_cond, gooding}. However, within conserving MBA:s,  
correlation functions may violate important properties such as positiveness \cite{pva_cond}. 

In spite of these possible drawbacks, the KBE, together with a MBPT approach to the self-energy,
offer the great advantage of incorporating non-locality and memory effects on equal footing.
For this reason they will be used in the next Section to benchmark our TDDFT results.
\section{Scrutinizing the linear response formalism  and the ALDA in lattice models}\label{RESULTS}
This section presents original results for three different topics. 
The first one is an illustration, by means of a simple model system, of the problems
one may encounter when using the LSS within the linear response formalism of TDDFT.

We then present two applications of lattice TDDFT, which offer a clear illustration of the limitations
(but also of  the positive aspects) of the ALDA. In the first case, we study the density response
function for the 3D Hubbard model, and we use an $f_{xc}^{ALDA}$ obtained from DMFT (Sect. \ref{DFTHub}).
The functional form of the kernel $f_{xc}^{ALDA}$ is also discussed.
In the second example, we study the non-linear, TDDFT-ALDA response of a finite cluster 
to strong, time-dependent perturbations. The $v_{xc}$ used in the ALDA is also in this case 
obtained from DMFT. The TDDFT results will be 
compared to exact ones, and to those from the Kadanoff-Baym dynamics introduced in Sec. \ref{KBE}. 
\subsection{Inversion of the non-interacting response}
\begin{figure}[b]
\centering
\includegraphics[width=87mm]{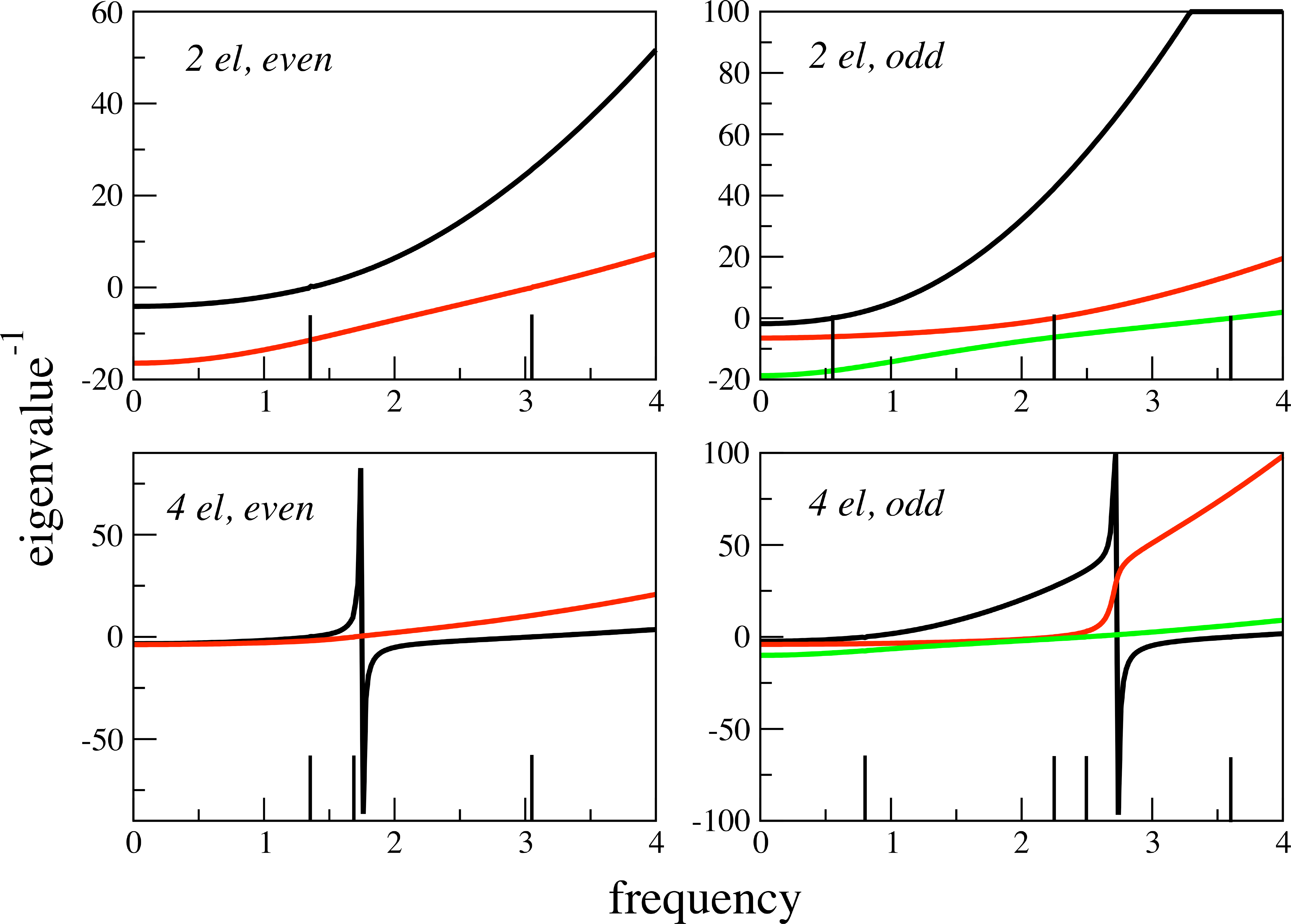}
\caption{(Color online) Inverse of the eigenvalues of $\chi_0$, for a six-site, spin-compensated chain with 2 and 4 electrons, as function of the frequency. The longer vertical ticks on the frequency axis denote
the excitation energies (see main text).}
\label{chi0_eigen}
\end{figure}
It was long hoped that the exact-exchange approximation within TDDFT would
provide a reasonable XC kernel which is both non-local and energy dependent and
thus constituting a significant step beyond the usual adiabatic approximations.
As mentioned in Sect. \ref{LinResp}, these hopes have recently evaporated, the main
reason being zeros in the non-interacting response function. This problem has 
actually been pointed some time ago \cite{MK}. In the spirit of the model investigations
at the heart of the present work we will here demonstrate the
problem in some model cases.

The non-interacting response function $\chi_0$ can always be written as 
\be
\chi_0(\vr,\vr';\w)  = \sum_{q} \frac{2 \w_{q} f_{q}(\vr) f_{q}^{*}(\vr')}{\w^2 - \w_{q}^2}
\ee
where the index $q$ is a double index $q =(k,k')$ and $\w_q = \e_{k'} - \e_k$ 
is an excitation energy given by the energy difference between an unoccupied ($\e_{k'}$)
and an occupied ($\e_k$) eigenvalue of the basic one-electron Hamiltonian.
The functions $f_q$ are excitation functions consisting of the corresponding products
between an unoccupied, $\vf_{k'}^{*}(\vr)$, and occupied, $\vf_{k}(\vr)$, orbital of the
same Hamiltonian. Due to the orthogonality of the orbitals $\{ \vf_k \} $, the functions $f_q$
always integrate to zero, resulting in the well known property of $\chi_0$ of a finite
system
\be
                            \int \chi_0(\vr,\vr') \;d^3r' =  0.
\ee
This means that the response function $\chi_0$ always has a zero eigenvalue at all
frequencies, which is a reflection of the physical fact that there can be no charge 
response to a constant potential. Unfortunately, the response $\chi_0$ also has 
other zero eigenvalues at specific frequencies. If we first consider a two-electron
problem, the lowest orbital $\vf_0$ is doubly occupied and all other orbitals are 
unoccupied. The complete set of orbitals {$\vf_k$} are, of course,
linearly independent, a fact which is not altered by removing the lowest member $\vf_0$
from the set. Neither is this fact altered by multiplying all the remaining orbitals in 
the set by the same function $\vf_0$. Consequently, in this case, all the excitation 
functions are linearly independent. Assuming for a moment that $\chi_0$ has an 
eigenvector $u(\vr)$ with a vanishing eigenvalue we find that    
\be 
\sum_{q} \frac{2 \w_{q} f_{q}(\vr) u_q}{\w^2 - \w_{q}^2} = 0, \label{vanish}
\ee
where the coefficients $u_q$ are the projections of the eigenfunction $u$ onto the 
excitation functions $f_q$. But, since the excitation functions are linearly independent,
all coefficients in this linear combination of the $f_q$:s must vanish and, since the 
frequency dependent factors are non-zero, we are led to the conclusion that all
the coefficients $u_q$ must vanish. But, again, the only function $u$ which is 
orthogonal to all the excitation functions $f_q$ is a constant and we have recovered
the already known case. Thus, in the two-electron case, there can be no additional
zero eigenvalues of $\chi_0$.

Proceeding now to more than two electrons, we have several sets of excitation functions,
one for each occupied orbital. Although all functions within a particular set are
linearly independent among themselves, we find it highly unlikely that the conjunction
of all sets are linearly independent. Indeed, every set is almost a complete set by
itself. One would therefore expect to be able to expand a member of one set in the
combined functions of two other sets. Consequently, by varying the frequency and the
coefficients $u_q$ in the vanishing sum above, Eq.(\ref{vanish}), one would expect to be
able to arrive at a particular vanishing linear combination of the excitation 
functions $f_q$. In the spirit of the present paper, we will illustrate these points
on a simple non-interacting linear chain of atoms with one orbital per site, no 
one-site energies, and only nearest neighbour hopping-  the same for each spin channel.

Choosing four atoms in the chain and two electrons with opposite spin, the relevant Hamiltonian is 4x4 and we have
only three excitation functions with corresponding excitation energies - from the
ground state to each of the three unoccupied states. As it turns out and is easily
verified on the back of an envelope, the three excitation functions in a 
four-dimensional space are definitely linearly independent and there are thus no
additional zero eigenvalues of $\chi_0$. 

Turning then to four electrons, two for each spin channel, we obtain four excitation functions from each of the
two occupied states to each of the remaining two unoccupied states. As easily verified,
two excitation functions are identical and so are their corresponding excitation energies.
In the sum above, Eq.(\ref{vanish}, these two functions will contribute equally and both are included
by multiplying one of the terms by two. Any linear chain has, of course, mirror symmetry
around the mid point, a fact which substantially reduces the necessary algebra -
especially for the few-site cases. The response function $\chi_0$ splits into a sum of
an even and an odd contribution and so do all the excitation functions. In the present
case, the two even excitation functions are identical and, consequently, the even part
of $\chi$ consits of only one term which cannot be orthogonal to anything but the constant
vector. The two odd excitation functions are easily seen to be linearly independent by
inspection. Thus, also in the four-electron case, there are no extra zero eigenvalues of
$\chi_0$ as long as there are four sites.

Including six sites in the model, we immediately approach the above discussed general
behaviour of $\chi_0$. Looking first at the two-electron case, we obtain five excitation
functions and energies. The calculations now become a little too complicated to carry
out by hand but are still rather trivial. The six-by-six determinant of the five 
excitation functions plus the constant vector turns out to be non-singular, thus 
demonstrating all excitation functions to be linearly independent. Hence, there are
no additional vanishing eigenvalues of $\chi_0$, in keeping with the general discussion above.

Having instead four electrons on the six sites gives us eight excitation functions of
which two turn out to be identical. But there is no way the remaining seven can be
linearly independent in a six-dimensional space. As discussed above, we would then
expect that a particular choice of frequency could result in coefficients giving rise
to a vanishing linear combination of excitation functions and thus a zero eigenvalue
of $\chi_0$. In Fig. \ref{chi0_eigen}  we have plotted the inverse of the five non-trivial frequency
dependent eigenvalues of $\chi_0$ for the even and the odd channels - for both the
two-electron case and the four-electron case. We see that these inverse eigenvalues
are all nice smooth functions of the frequency in the former case. We also see that
two eigenvalues in the four-electron case pass zero at particular frequencies 
away from the excitation energies - one in the odd channel and one in the
even channel. These results corroborate the correctness of the general discussion 
above.     
\subsection{Linear response in the ALDA for the 3D Hubbard model.}
\begin{figure}[t]
\centering
\includegraphics[width=75mm]{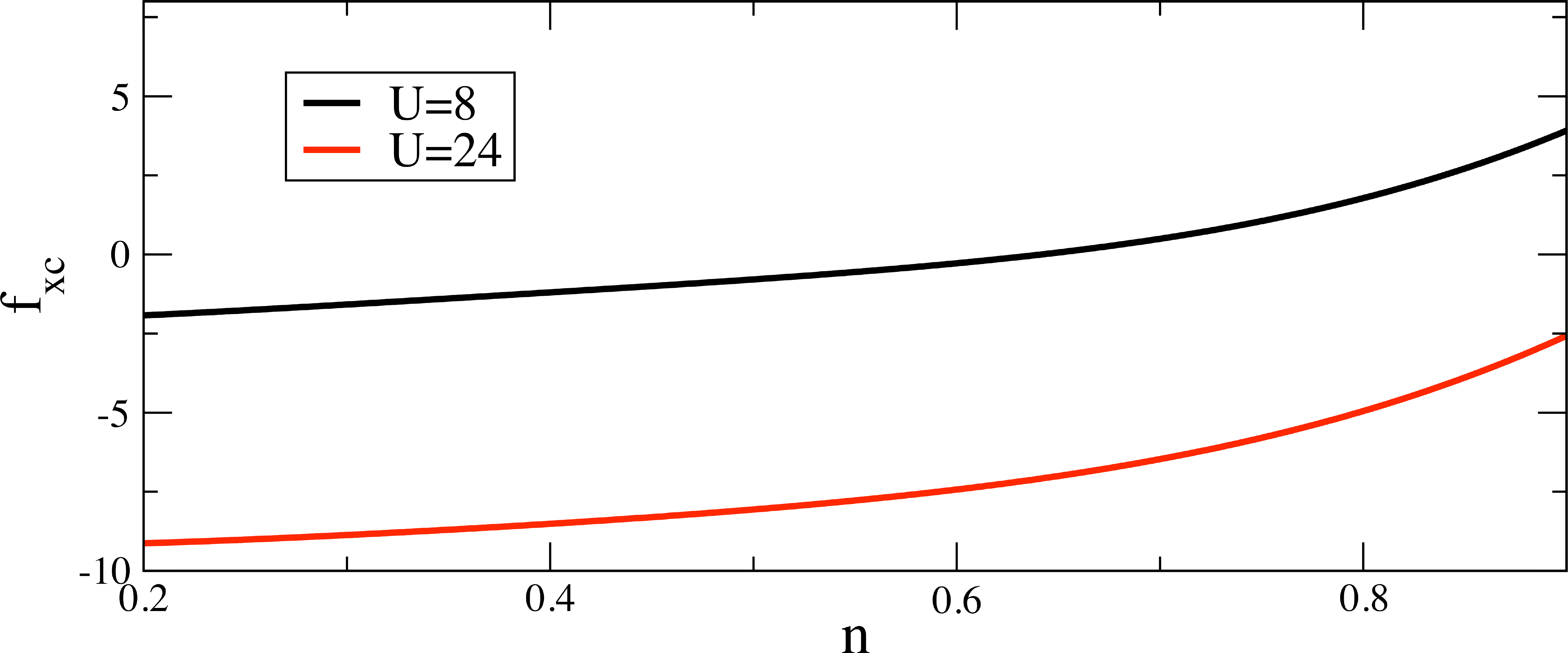}
\caption{(Color online) The ALDA exchange-correlation kernel $f^{ALDA}_{xc}$ for $U=8$ and $U=24$ 
obtained by differentiating $v_{xc}$ with respect to the density.}
\label{fxcfig}
\end{figure}
In analogy to the continuum case, the general expression for the XC kernel  $f_{xc}$ on the lattice is
\begin{equation}
f_{xc}({\bf R},{\bf R'};t,t')=\frac{\delta v_{xc}({\bf R},t)}{\delta n_{\bf R'}(t')}.
\end{equation}
In the ALDA, an ordinary derivative in the density is performed, and the space and time dependence
is entirely local:
\begin{align}
f_{xc}^{ALDA}({\bf R},{\bf R'};t,t')&=\frac{d v_{xc}(n_{\bf R}(t))}{dn_{\bf R'}(t')}\nonumber\\
&= v^{\prime}_{xc}(n_{\bf R}(t))\delta_{\bf RR'}\delta(t-t'),
\end{align}
and $f_{xc} \equiv f_{xc}^{ALDA}({\bf q},\omega)$, i.e. $f_{xc}$ only depends of the (uniform) ground-state density
of the system.
In the Hubbard model of Eq. (\ref{Hamiltonian}), the bare interaction $U$ is also local in time and space. 
Accordingly, for the 3D homogeneous Hubbard model, and in the (${\bf q},\omega $) space, 
the ALDA density-density response function $\chi (\textbf{q},\omega)$ is 
\begin{align}
 \chi (\textbf{q},\omega) = \frac{\chi _{0} (\textbf{q},\omega)}{1-(U+f_{xc})\chi_{0} (\textbf{q},\omega)}\label{chiDyson}
\end{align}
Here, $\chi (\textbf{q},\omega)$ is the Fourier transform of the retarded response function
\begin{align}
 \chi ({\bf R},t) = -i\theta(t) \langle  \left[ \tilde{n}_{\bf R}(t),\tilde{n}_{\bf 0}(0) \right ] \rangle _{gs},
\label{chi:definition}
\end{align}
and $\tilde{n}_{\bf R}(t)=\hat{n}_{\bf R}(t)-\langle \hat{n}_{\bf R}(t)\rangle$, with $\hat{n}_{\bf R}(t)=\hat{n}_{\bf R \uparrow}(t)+\hat{n}_{\bf R \downarrow}(t)$.
Furthermore, $\chi_{0}$ is the response function of the Kohn-Sham system,
\begin{align}
 \chi_{0} (\textbf{q},\omega) = \frac{2}{(2\pi)^3} \int d^3 k \frac{n_F(\epsilon_{\textbf{k}}) - n_F(\epsilon_{\textbf{k}+\textbf{q}})}{\epsilon_{\textbf{k}} - \epsilon_{\textbf{k}+\textbf{q}} + \omega + i\eta},
\end{align}
where  $\epsilon_{\bf k}=-2t(\cos k_x + \cos k_y + \cos k_z)$ is the single particle energy dispersion 
in the 3D simple cubic lattice, $n_F(\epsilon)$ is the Fermi function (we work at zero temperature) 
and the integral in ${\bf k}$ is performed
over the first Brillouin zone.\newline

Before presenting the random-phase-approximation (RPA) and TDDFT-ALDA results for $\chi$, it is useful
to discuss briefly the features of $f_{xc}$ in the ALDA. In Fig. \ref{fxcfig}, we show $f^{ALDA}_{xc}$ as a function
of the density $n$. Since $f_{xc}^{ALDA}(n)=f_{xc}^{ALDA}(2-n)$, we can consider results for $n\leq 1$.
Furthermore, the results $f_{xc}$ for $n\leq 0.2$ and $n \ge 0.95$ exhibit significant noise, and thus are not displayed.
In Fig. \ref{fxcfig}, we see clearly one of the interesting features of $f_{xc}^{ALDA}$, namely the XC kernel can 
be either negative or positive \cite{also1D}. This is especially evident for $U=8$, and can also immediately be gathered from
the results for $v_{xc}$ in Fig. \ref{vxcDMFT}. Thus we see that, depending on the band filling, $f_{xc}^{ALDA}$
can reduce or reinforce the effect of the bare interaction $U$ in Eq. (\ref{chiDyson}). This is at variance with
the usual case, where the XC kernel tend always to induce an effective interaction which is reduced with respect
to the bare one.
The other important aspect in the XC kernel is its behavior at $n=1$. Looking again at Fig. \ref{vxcDMFT},
we note that the discontinuity in $v^{DMFT}_{xc}$ for $U > U^{Mott}_{c}$ (in our case $U=24$) will introduce 
a (Dirac's delta-like) spike in the XC kernel at $n=1$. This singular behavior would be absent
for $U < U^{Mott}_{c}$ in 3D, but always present in 1D where $v_{xc}$ is discontinuous for any value of $U$.
\begin{figure}[t]
 \centering
 \includegraphics[width=85mm]{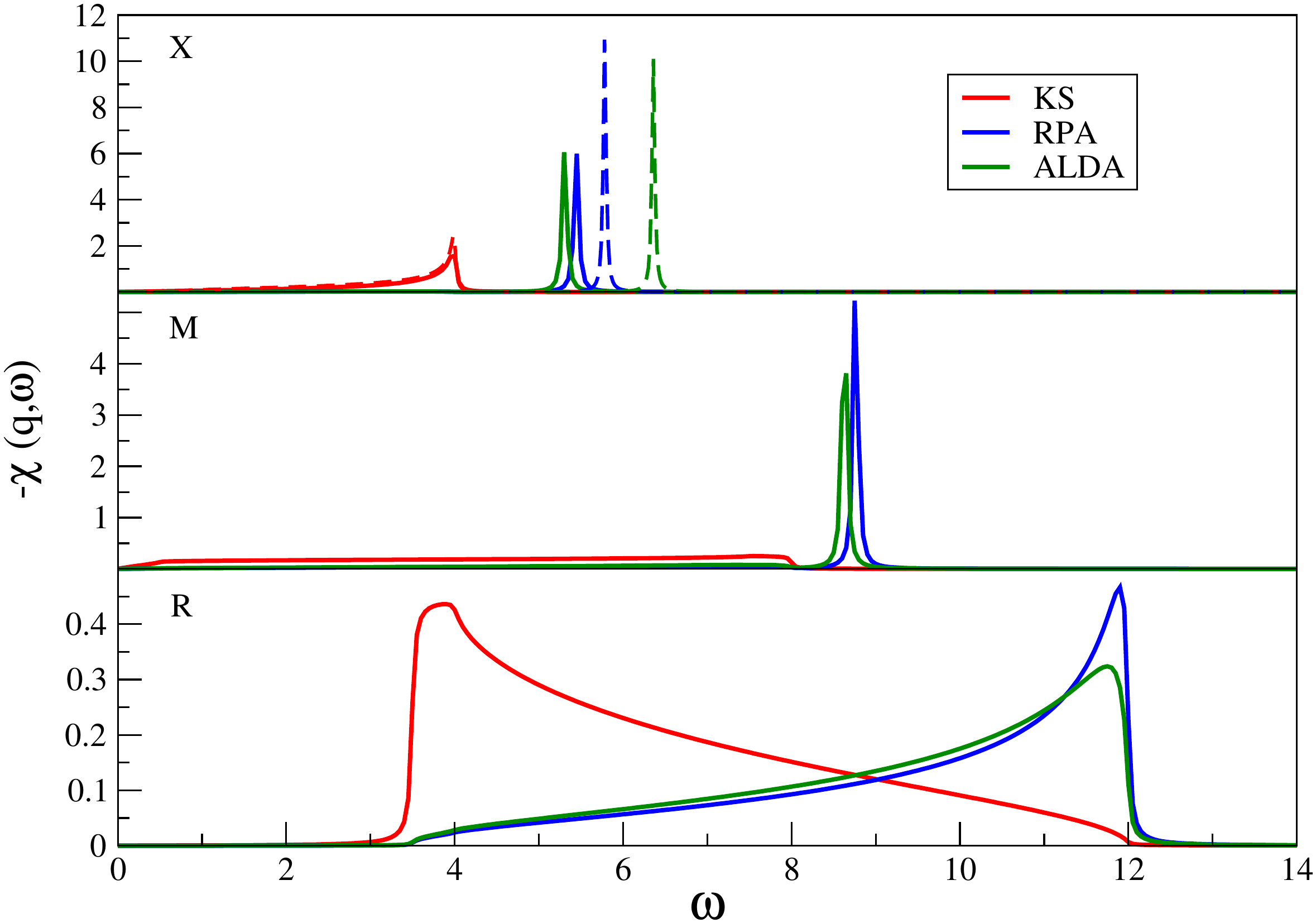}
\caption{(Color online) The imaginary part of $-\chi (\textbf{q},\omega)$ for $U=8$, as a function of $\omega$, at 
three high symmetry points in the Brillouin zone. The chosen ${\bf q}$ values are $X = (\pi,0,0)$, $M = (\pi,\pi,0)$ 
and $R = (\pi,\pi,\pi)$ (note the different vertical scale for the three panels). The average density is $n=0.5$, i.e. quarter 
filling. In each panel, the solid curves refer to the  Kohn-Sham,  RPA  and ALDA cases (red, blue and green, 
respectively). The effect of $f_{xc}$ for this filling is to give a reduced effective interaction, with the ALDA peak 
shifted to the left of the RPA one. For the point $X$ we also show results for another filling ($n = 0.85$, dashed curves), 
where $f_{xc}$ is positive. In this case, the ALDA peak (green curve) is at the highest energy.}
\label{chiqom}
\end{figure}

Results for $\chi({\bf q}, \omega)$ for three values of ${\bf q}$ are shown in the three panels of Fig. \ref{chiqom}. 
The systems density that we consider is quarter filling, i.e. $n=0.5$, and $U=8$.
In each panel, there are three solid curves which represent  the imaginary part  of the Kohn-Sham (red curve), RPA (blue curve) and
TDDFT-ALDA versions of  $-\chi({\bf q}, \omega)$ (the dashed curves in the top panel will be discussed momentarily).
For the RPA and ALDA curves, we see sharp structures, outside the $\chi_{0}$ continuum region, 
corresponding to the zeros of  $1-(U+f_{xc})\chi_{0} (\textbf{q},\omega)$. Such peaks describe the plasmonic features
of our system. We mention in passing that we have performed several test of our numerics, including a (successful) verification
the of the f-sum rule, 
\begin{align}
 &M(\textbf{q}) = -\frac{2}{\pi} \int \text{Im} (\chi_0(\textbf{q},\omega)) \omega d\omega=\nonumber\\
 & -\frac{2}{(2 \pi)^3} \int d^3k \langle c^\dagger _k c_k \rangle (\epsilon (\textbf{k}+\textbf{q}) + \epsilon (\textbf{k}-\textbf{q}) -2\epsilon(\textbf{k})).
\end{align}
It interesting to note that, for the solid curves, the RPA peaks are always at higher energy than the ALDA ones. This is easily
explained considering that at quarter filling, and for $U=8$, $f^{ALDA}_{xc}$ is negative (Fig. \ref{fxcfig}), thus screening the bare $U$. 
However, it also clear that for densities close to half-filling, say $n=0.85$, the XC kernel at $U=8$ is positive, giving ALDA peaks 
at energies higher than the RPA ones. This situation is shown in the top panel of Fig. \ref{chiqom}, for one value of $\bf q$, 
and it corresponds to the dashed curves in that panel. %

\begin{figure}[htb]
 \centering
 \includegraphics[width=85mm]{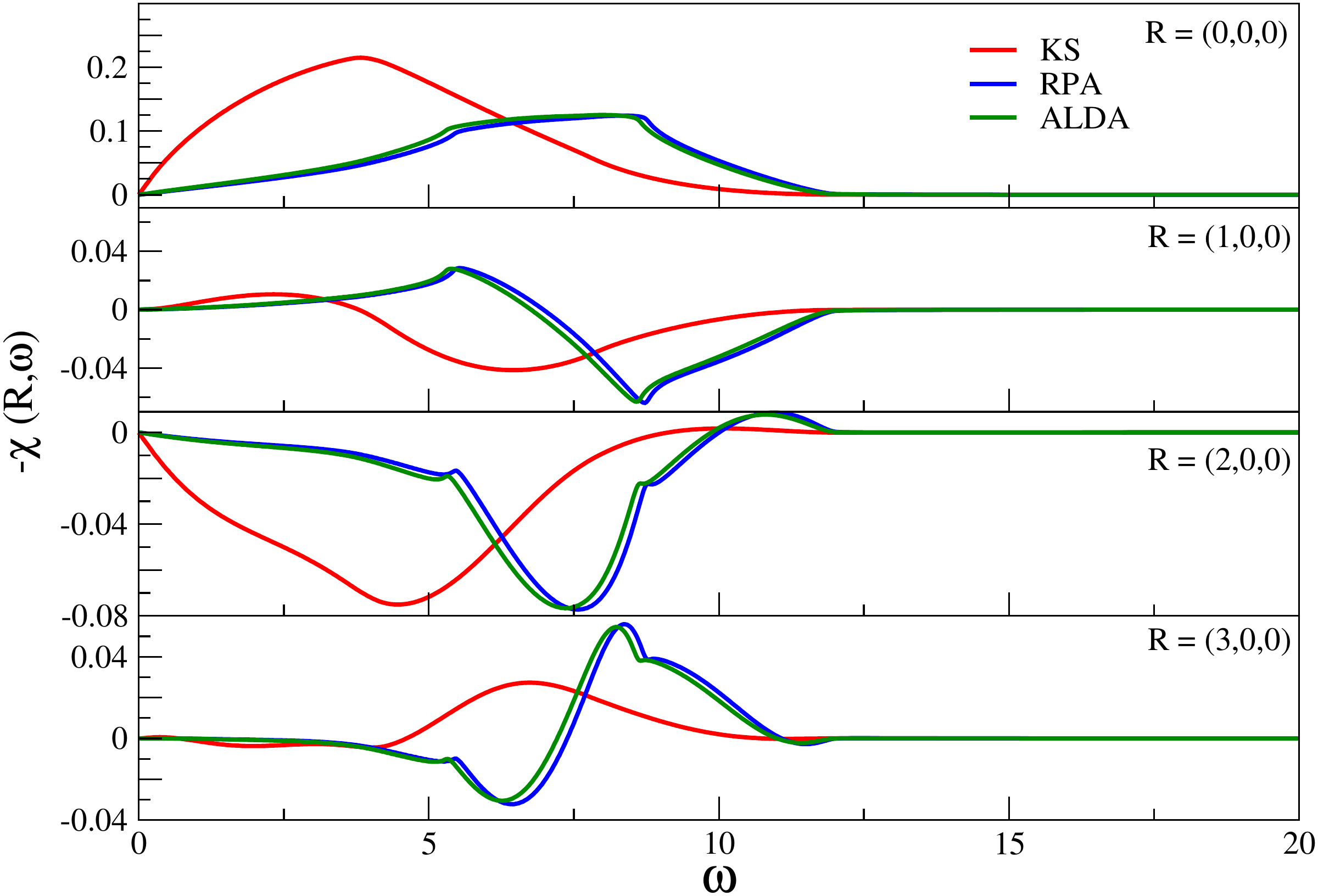}
\caption{(Color online) The linear response $\chi (\textbf{R},\omega)$ as a function of $\omega$ for $U=8$, 
for different values of $\textbf{R}$.}
\label{chi_realspace}
\end{figure}

A similar behavior can be observed in real space. In Fig. \ref{chi_realspace}, which refers to the case of quarter filling and $U=8$,
the four panels show the dependence of $\text{Im} \chi({\bf R}, \omega)$ on the distance $|{\bf R}|$ along the $x$-axis
(as  one goes away from $R=0$, the response function diminishes quite rapidly). 
The principal effect of the correlations is to move the structures in $\text{Im} \chi$ to higher energies with respect of the 
Kohn-Sham results. Furthermore, we can notice how the ALDA profiles are always shifted at lower energies with 
respect to those of the  RPA, due to the negative value of $f_{xc}^{ALDA}$ when $n=0.5$ and $U=8$.

In general, the pair correlation function (and thus the density response function) is a careful indicator of the accuracy of a 
many-body approximation.  Thus, looking at properties which involve $\chi$ can provide a severe test to
assess the performance of the ALDA. For example, the density-density response function can be used to compute the 
ground-state energy of the system. More specifically, via the connection
to the dynamical structure factor $S(\bf{q},\omega)$, one can determine the local double occupancy 
$d_{\bf R}=\langle \hat{n}_{\bf R \uparrow} \hat{n}_{{\bf R} \downarrow} \rangle$. 
We have used $\chi$ in the RPA and the ALDA to compute $d_{\bf R}$, and the results
are reported in Table \ref{occups}. Two general features can 
be noted. The first is that in some cases the RPA and ALDA results for $d_{\bf R}$ are negative. Since $d_{\bf R}$ is manifestly positive, this is an adverse 
outcome of the ALDA, and is rather general in character. Indeed, it is a long established fact that, in the RPA \cite{Lindhard},
pair correlations for the electron gas at metallic densities are strongly negative 
at short distances \cite{Alf}; more recently, the same problem has 
been also experienced within the self-consistent GWA and BA \cite{Holm,Dimercondmat}. 

The second general trait in Table \ref{occups} is related to the sign of $f_{xc}$: at quarter filling, where the $f_{xc}$ 
is negative, we get an effective lower interaction, which tends to increase the ALDA double occupancy with respect to 
the one from  the RPA. On the other hand, at $n=0.85$, the $f_{xc}$ for $U=8$ is positive, and thus instead increases 
the interaction, which results in a lower double occupancy. 

As a conclusive remark, while our results for the double occupancy point out a shortcoming of the ALDA, they also
indicate that the problem with the sign of $d_{\bf R}$ could be less acute than in the continuum case.
In fact, at least for the cases we considered, double occupancies  in the ALDA are only slightly negative and,
most likely, this  is a consequence of the natural cut-off introduced by the lattice at short distances.

\begin{table}
\begin{center}
    \begin{tabular}{ | l | l |l|l|l|}
    \hline
     Parameters &$d_{DMFT}$& 
    $d_{KS}$ &
    $d_{RPA}$ &
    $d_{ALDA}$  \\ \hline
    $U=8, n=0.5$ &0.036& 0.062 &  −0.010 & −0.007 \\ \hline
    $U=24, n=0.5$ &0.016& 0.062 & −0.047 & −0.035 \\ \hline
    $U=8, n=0.85$ &0.114& 0.178 & +0.072 & +0.059 \\ \hline
    \end{tabular}
\end{center}
\caption{Double occupancies $d=\langle n_\uparrow n_\downarrow \rangle$ obtained from DMFT, compared against results from linear response. 
Results from $\chi _{0}$, $\chi _{RPA}$ and from $\chi _{ALDA}$ are shown.}
\label{occups}
\end{table}
 
%
\subsection{Exact, ALDA and Kadanoff-Baym real-time dynamics in a small cluster.} 
\begin{figure}[b]
\includegraphics[width=70mm,angle=-0]{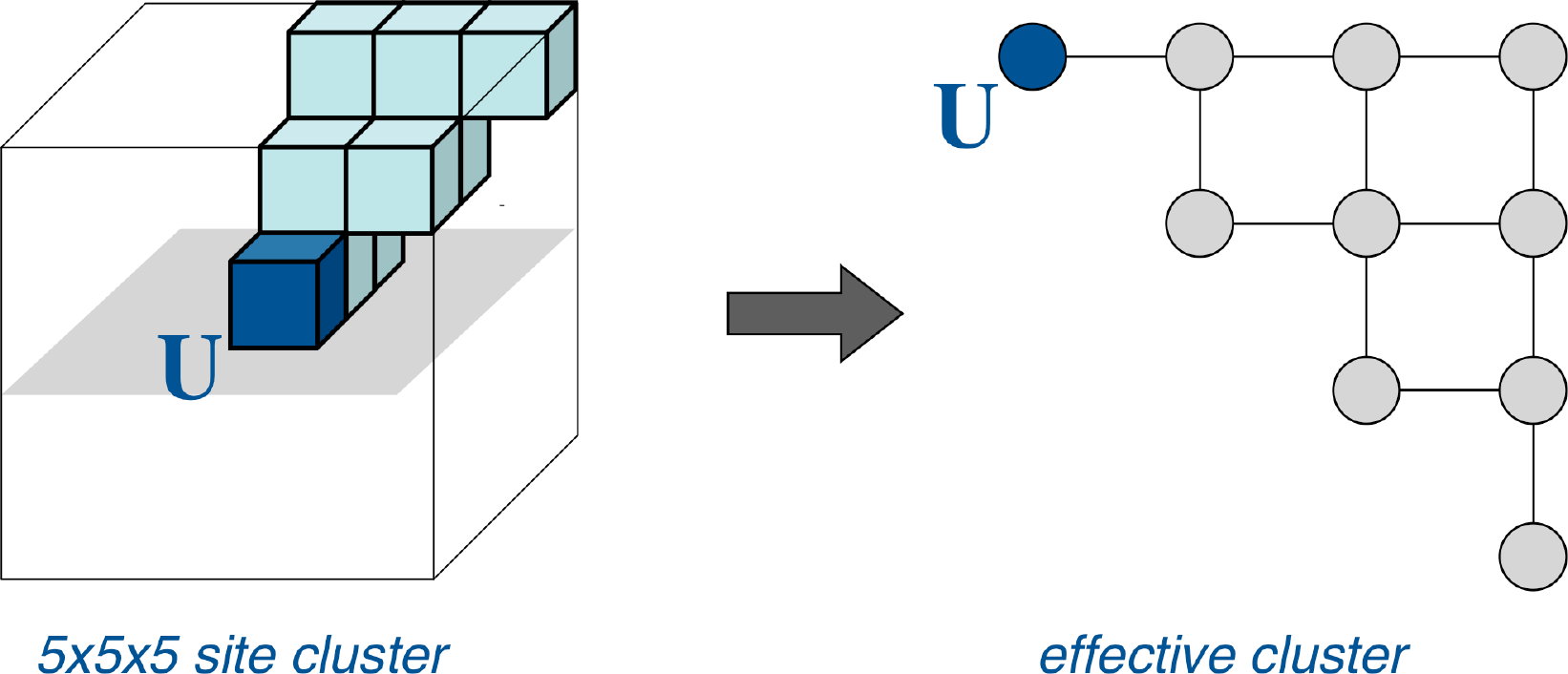}
\caption{(Color online). Original 125-site cluster (left) and its effective image (right). In the original cluster, all sites
are related by symmetry to the 10 sites explicitly shown. Accordingly, the site $0$ is the same in the two clusters.}
\label{Clusters} 
\end{figure}
The system we choose to compare TDDFT-ALDA, exact, and Kadanoff-Baym results is 
a simple cubic cluster with $5^3$ sites and open boundary conditions (Fig. \ref{Clusters}, left). 
We choose the cluster to be highly inhomogeneous, by having 
a single interacting impurity in the center $R=0$, {\it e.g.}, $U_R=U\delta_{R0}$ in Eq. (\ref{Hamiltonian}). 
We also set $w_R(\tau)=w_0(\tau)\delta_{R0}$.
Due to cubic symmetry, only $N_{sym}=10$ out of $125$ one-particle eigenstates, those with 
non-zero  amplitude at $R=0$, determine
the static and time-dependent density at $R=0$, making the size of the exact configuration space
manageable \cite{MCCV87}, in the guise of an effective 10-site cluster (Fig. \ref{Clusters}, right).
Furthermore, if $\epsilon_{R\neq 0}=0$, further use of symmetry gives $N_{sym}=7$. In this case, the shape
of the cluster corresponds to the $R=0$ impurity directly connected with to all the other 6 sites.
However the $R\neq0$ states are not connected directly to each other, but only to the impurity. 
In the ground state, the way $N_e$ electrons in the cluster distribute between the $N_{sym}$ active and 
$125-N_{sym}$ spectator states corresponds to the exact many-body eigenstate with lowest energy. 
As a overall remark, we note that the simplification due to symmetry is very peculiar to our choice of a 
local interaction at a single site, and permits us to obtain exact dynamical results in 3D for a quite large cluster. 
Furthermore, such simplification remains in the presence of a time-dependent perturbation localized at the impurity site.

\begin{figure}[t]
 \centering
 \includegraphics[width=90mm]{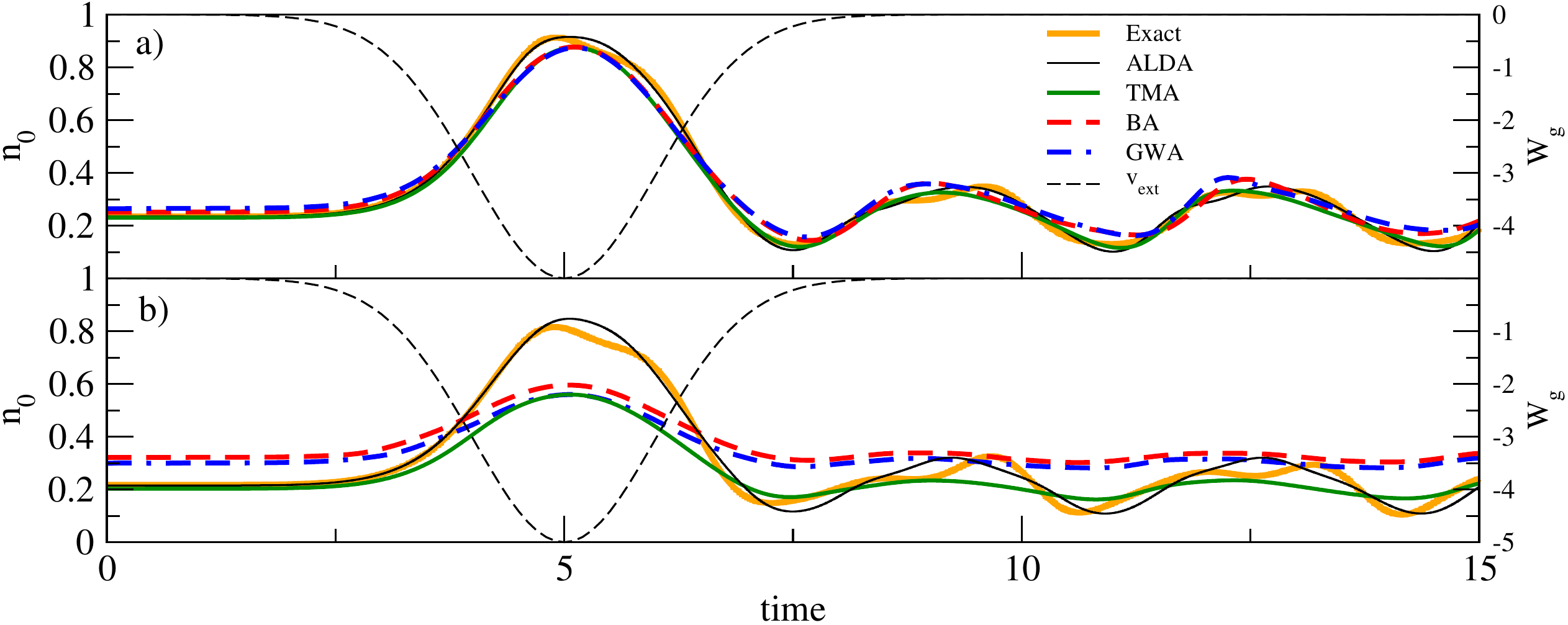}
\caption{(Color online) Time-dependent density, $n_0(\tau)$, of the central (interacting) site in a 
5x5x5 simple cubic cluster subject to a Gaussian potential $W_g(\tau)=-5e^{-\frac{\left(\tau-5\right)^2}{2}}$ (thin black dashed). 
The curves correspond to exact (thick solid orange), ALDA (thin solid black), 
BA (red dashed), GWA (blue dot-dashed) and TMA (solid green). 
The number of particles per spin of the equivalent 
cluster is 2. In a) $U=8$ and in b) $U=24$.}
\label{KBE_vs_exact1}
\end{figure}
In Fig. \ref{KBE_vs_exact1} we study the time-dependent density of the 
interacting impurity when subject to a slowly varying pulse for 
different values of the interaction strength.  This is a situation in which the
ALDA is expected to be appropriate. In the ground state, before the 
perturbation has been introduced, the systems are in the low density
regime. The reason of this choice is that  of the three MBA:s we consider,
one of them, the TMA, performs rather well at these densities \cite{pva1},
while the BA and GWA are not good in any of the regimes we considered.

In Fig. \ref{KBE_vs_exact1}, where the interaction strength
is weak and the external field is slowly varying, we see that the ALDA,
as well as the KBE+MBPT, give an overall good description, both in 
the time window where the perturbation is actually present,
and afterwards, when the system is performing free oscillations.
When the interaction strength is geared up in Fig. \ref{KBE_vs_exact1}b 
we clearly see that the ALDA is superior to all the KBE results from the MBA:s,
while the TMA performs better than the BA and the GWA. It is fair to
say that overall ALDA provides a good description for these slow external fields.

\begin{figure}[t]
\centering
\includegraphics[width=85mm]{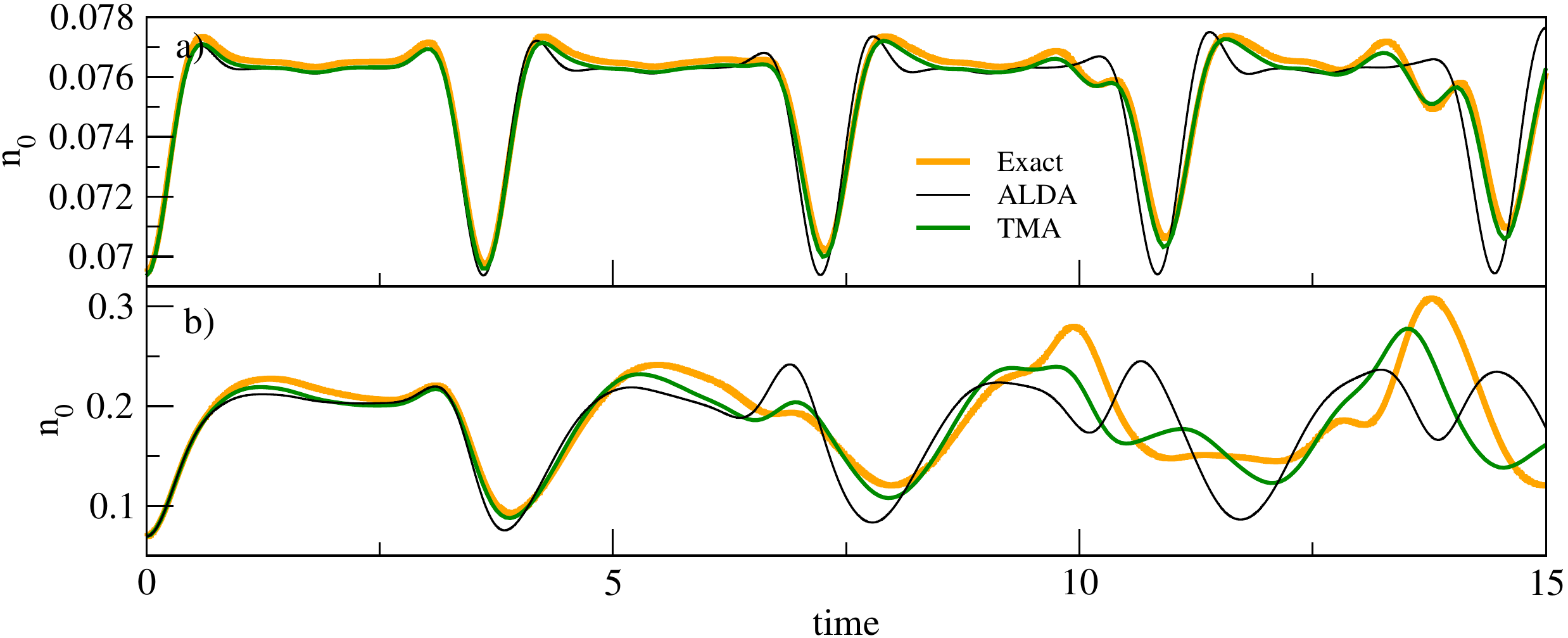}
\caption{(Color online) Time-dependent density, $n_0(\tau)$, of the central (interacting) site in a 
5x5x5 simple cubic cluster subject to a step potential, $W_g(\tau)=W_0\Theta(\tau)$. 
The curves correspond to exact (thick orange), ALDA (thin black) and 
TMA (green). The number of particles per spin of the equivalent 
cluster is 1. In a) $W_0=-0.2$ and in b) $W_0=-2$.}
\label{KBE_vs_exact2}
\end{figure}

In Fig. \ref{KBE_vs_exact2}, we show the exact, ALDA, and TMA results for case of a moderate interaction 
strength and a step perturbation (BA and GWA, not presented, perform quite poorly). In this case, 
the shortcomings of the ALDA are evident. For example, the ALDA does not reproduce the long-time 
oscillations properly: in Fig. \ref{KBE_vs_exact2}a, the ALDA performs well for $\tau \lesssim 4$, but after that, 
ALDA suffers an increasing dephasing with respect to the exact and TMA solutions. On the other hand, the 
TMA and the exact curve are in very good mutual agreement for this situation,
(see {\it e.g.} the region $\tau \approx14$), which corresponds to a highly non-adiabatic, but weak, perturbation.

For a stronger step perturbation  (Fig. \ref{KBE_vs_exact2}b), the KBE+MBPT approach, represented here 
by the TMA, is able to describe the history dependence quite well, and 
significantly better than the ALDA, which becomes unreliable. However, when both the interaction is strong and the external 
field is highly non-adiabatic, both the ALDA and the KBE+MBPT fail to describe the transient dynamics 
(not shown).

To summarize,  the ALDA, in suitable conditions, i.e. for slow perturbations,
can perform well. However, we wish to remark that the
ALDA we have considered here accounts for the effect of correlations in a non perturbative way, since it is
obtained from DMFT reference calculations. Otherwise, for systems with much stronger interactions,
a failure of ALDA should be expected also for external fields which vary slowly in time.
\section{Conclusions}\label{CONCLUDE}
The application of time-dependent density-functional theories to the non-equilibrium dynamics
of lattice models has recently received some attention in the scientific community. Conceptually firm ground has now been 
established for approaches which use the current density (or, more properly, the lattice bond current) as basic variable. 
By contrast, the early TDDFT-like approaches for lattice systems made use of the  
density as the basic variable. On the lattice, a density-based approach lacks a completely rigorous formulation, but it appears possible to 
consider it on heuristic grounds, in the sense that the missing rigor should not be of significant practical consequence.

It is in this spirit  that in this work we have adopted lattice TDDFT as an investigative tool to discuss two general
aspects of TDDFT, which are present also in the continuum formulation. These are the linear response 
formalism in TDDFT, and the adiabatic local density approximations (ALDA). 

Instrumental to this strategy, the first part of our paper has been devoted to short reviews of linear response theory, 
lattice TDDFT, and also of the non-equilibrium  Green's function method: the latter method was used
to benchmark the results from lattice TDDFT.

In the second part of this contribution, where our new results are presented, we have studied the linear and non-linear response
of 1D and 3D lattice Hubbard-type models. We find that the limitations encountered in the continuum case by
the TDDFT linear response persist in simple 1D lattice model systems. Furthermore, an ALDA treatment  for the linear response
of the 3D homogenous Hubbard model exhibits an important pitfall which is also present in the continuum case, namely short-range 
correlations are incorrectly accounted for by the ALDA, albeit at a lesser extent than for the electron gas. 

We have also shown 
that the sign of the ALDA XC kernel can change depending on the values of $U$ and the density, and that for large enough
values of $U$, $f_{xc}$ can exhibit a singular behavior at half-filling (due to the discontinuity in $v_{xc}$, which in turn is indicative of the Mott-Hubbard
metal-insulator transition). Such characteristics of $f_{xc}$ are clearly visible in the energy dependence
of the density response function. To our knowledge, the qualitative features of $f^{ALDA}_{xc}$ just mentioned have no correspondence in the available  XC kernels for continuum case. On general grounds, we expect they should be robust against the 
inclusion of non-local, non-adiabatic effects. 

There have been attempts to go beyond the ALDA and simultaneously include non-locality and frequency dependence in the kernel $f_{xc}$.
Several of these attempts have originated in the LSS or, equivalently, in the variational approach starting from the Klein functional, and  have,
therefore, led to severe difficulties as discussed in the introduction. 
The problem is associated with use of
the LSS or, within the variational appoach, with starting from the Klein
functional. The "perpetrator" is the double inversion of the $\chi _{0}$
within these formulations. We can already anticipate that this particular
problem will not be present if one instead choses to start from the more
sophisticated Luttinger and Ward (LW) functional \cite{LW}. 
Whether or not the LW functional will give rise to other problems remains
to be seen. Work along these lines is underway.

In the non-linear case, our results can be summarized as follows. An ALDA which accounts for the effect of correlations in a 
non-perturbative way, can provide a satisfactory description of the non-equilibrium dynamics induced by slow perturbations.
When non-local and non-adiabatic effects play an important role, i.e. for fast perturbations, the ALDA is clearly insufficient.
This is especially evident if one compares the exact, TDDFT-ALDA and Kadanoff-Baym dynamics.  In the latter, effects
beyond the ALDA are present, and are seen to be necessary to get a good agreement with the exact data.

In conclusion, both in the linear and non linear cases, our results for lattice models confirm that going beyond the ALDA is
certainly necessary in many instances, and that lattice models can be a useful way to scrutinize fundamental open issues
in TDDFT and to explore possible avenues for improvement. 
\section*{Acknowledgments}
It is a pleasure to acknowledge Antonio Privitera, who is currently
visiting our Division, for useful discussions about DMFT. CV
wishes also to acknowledge discussions with Klaus Capelle and Gianluca Stefanucci.
This work was supported by ETSF (INFRA-2007-211956).
%


\begin{thebibliography}{92}
\bibitem{t_meaning}
The three terms in the RHS of Eq. (\ref{Hamiltonian}) represent the kinetic, interaction, and perturbation contributions to $H$.
In particular, $t$ denotes, as usual, the hopping term; later on in the paper, Sect. \ref{KBE}, the symbol $t$ will have a different meaning, 
namely a time label on the Keldysh contour.
\bibitem{definitions} Here $\omega_n=(2n+1)\pi/\beta$, $\beta^{-1}=T$ ($k_B=1$).
 To obtain the ground state energy, we let $\beta \to \infty$.
\bibitem{notation} Matrix notation/multiplication in space and spin indexes is adopted.
\bibitem{HF}We do not consider the Hartree-Fock approximation, since it
includes no correlation effects.
\bibitem{also1D} A similar behavior is also found in the 1D case.
\end{thebibliography}
\end{document}